# A Constrained, Weighted-$\ell_1$ Minimization Approach for Joint Discovery of Heterogeneous Neural Connectivity Graphs

**Chandan Singh**
University of California, Berkeley

**Beilun Wang**
University of Virginia

**Yanjun Qi**
University of Virginia

## Abstract

Determining functional brain connectivity is crucial to understanding the brain and neural differences underlying disorders such as autism. Recent studies have used Gaussian graphical models to learn brain connectivity via statistical dependencies across brain regions from neuroimaging. However, previous studies often fail to properly incorporate priors tailored to neuroscience, such as preferring shorter connections. To remedy this problem, the paper here introduces a novel, weighted-$\ell_1$, multi-task graphical model (W-SIMULE). This model elegantly incorporates a flexible prior, along with a parallelizable formulation. Additionally, W-SIMULE extends the often-used Gaussian assumption, leading to considerable performance increases. Here, applications to fMRI data show that W-SIMULE succeeds in determining functional connectivity in terms of (1) log-likelihood, (2) finding edges that differentiate groups, and (3) classifying different groups based on their connectivity, achieving 58.6% accuracy on the ABIDE dataset. Having established W-SIMULE's effectiveness, it links four key areas to autism, all of which are consistent with the literature. Due to its elegant domain adaptivity, W-SIMULE can be readily applied to various data types to effectively estimate connectivity.

## 1 Introduction

Recently, there has been great interest in mapping the interactions between brain regions, a field known as functional connectomics (Smith et al. 2013b). The resulting maps, or connectomes, are fundamental to the study of neuroscience, as having a map of the brain allows for understanding neural pathways and systems (Seung 2011). Furthermore, these connectomes have immediate applications to pathologists trying to understand the neural characteristics underlying clinical disorders (Uddin et al. 2013).

Here, we focus on the important problem of estimating brain connectivity for more than one group (*i.e.* a disease group and a control group). Generally, studies use the simple pairwise correlations (*i.e.* the pearson correlation coefficient) between the activity of different areas as markers of a connection (Rogers et al. 2007). However, from a neuroscience perspective, this fails to extract the conditional correlations present in the brain and results in spurious connections.

Mathematically, determining functional connectivity amounts to first calculating a covariance matrix ($\Sigma$) from the data and then estimating the connectivity graph with the precision matrix ($\Omega = \Sigma^{-1}$). Zeros in $\Omega$ correspond to conditionally independent nodes, while non-zero values represent conditional edges (Lauritzen 1996). Recently, Gaussian graphical models (GGMs) have proven to be well-suited to estimating $\Omega$ (Koller et al. 2007).

This study's main contribution is the novel formulation of W-SIMULE, which arises naturally from brain-imaging data. W-SIMULE is a weighted-$\ell_1$, multi-task graphical model which robustly estimates $\Omega$ for each group. The main advantages of this method are:

- Effectiveness: it yields quantifiably accurate connectivity in terms of log-likelihood and classification accuracy
- Domain adaptivity: it elegantly enforces a prior based on the problem at hand and can overcome the often-incorrect Gaussian assumption by using nonparanormality
- Interpretability: it calculates a connectome for each group which can be tuned to the desired sparsity level and is particularly effective at low sparsity levels
- Efficiency: the formulation is column-wise parallelizable and quickly solvable

This study examines a large, resting-state fMRI dataset which serves to compare and validate several recent multi-task learning models. W-SIMULE outperforms other graphical models on this dataset in terms of (1) maximizing the log-likelihood of the connectome, (2) finding edges that differentiate groups, and (3) classifying subjects into their group (autism vs. control). Finally, W-SIMULE is used to analyze the neural basis of autism.

The organization of the paper is as follows: Sec. 2 reviews related work, Sec. 3 develops the model, Sec. 4 shows experiments demonstrating the effectiveness of W-SIMULE, and Sec. 5 explains the conclusions.

## 2 Background and Related Work

A variety of related work exists. We divide it into four categories: (1) weighted-$\ell_1$ models, (2) brain connectivity priors, (3) multi-task brain studies, and (4) multi-task baselines. None of the existing work meets all the specifications of W-SIMULE. Notably, W-SIMULE outperforms all previous approaches in terms of (1) effectiveness, (2) domain adaptivity, (3) interpretability, and (4) efficiency. Table 1 summarizes the related work. For an overview of ABIDE classification studies, see Appendix Table A1.

Table 1: Summary of related work.

| Method | Conditional Independence | Multi-task | Column-wise Parallelizable | Imposes Prior |
|---|---|---|---|---|
| W-SIMULE | ✓ | ✓ | ✓ | ✓ |
| CLIME | ✓ | ✓ | ✗ | ✗ |
| GLASSO | ✓ | ✓ | ✗ | ✗ |
| SIMULE | ✓ | ✓ | ✓ | ✗ |
| JGL | ✓ | ✓ | ✗ | ✗ |
| SIMONE | ✓ | ✓ | ✗ | ✗ |
| DPM | ✓ | ✓ | ✗ | ✗ |
| Spatial Regularization | ✗ | ✗ | ✗ | ✓ |
| Weighted-$\ell_1$ GGMs | ✓ | ✗ | ✗ | ✓ |
| sGGGM | ✓ | ✓ | ✗ | ✗ |
| MNS | ✓ | ✓ | ✗ | ✗ |

**Weighted-$\ell_1$ Models.** $\ell_1$ norms effectively induce sparsity in graphical models (Friedman, Hastie, and Tibshirani 2008). Importantly, by weighting the $\ell_1$ norm with a prior, the norm can induce sparsity while simultaneously penalizing the selection of certain edges[1]. Some recent studies use a weighted-$\ell_1$ norm to enforce a prior on a Gaussian graphical model. For example, one model uses reweighted-$\ell_1$ norms to maintain sparsity while reducing penalties on nodes with high degree, thus encouraging the appearance of "hub" nodes with a large number of connections (Liu and Ihler 2011). Here, spatial penalization does not necessarily give rise to hub nodes, but rather disincentivizes all nodes from making long connections. Another study uses weighted-$\ell_1$ optimization to improve neighborhood selection for gene network estimation (Shimamura et al. 2007). However, no previous weighted-$\ell_1$ study extends to multi-task learning or brain connectivity.

**Brain Connectivity Priors.** W-SIMULE requires choosing a prior to enforce. For fMRI data, spatial distance is a strong candidate, as spatially distant regions are less likely to be connected in the brain (Watts and Strogatz 1998; Vértes et al. 2012). Previous studies have utilized spatial regularization, but use it for smoothing rather than feature selection (Ng and Abugharbieh 2011; Grosenick et al. 2011). Notably, one recent study uses weighted-$\ell_2$ regularization to generate ROIs for brain connectivity (Baldassano et al. 2012). Another recent study uses a weighted prior to enhance a neighborhood selection algorithm (Bu and Lederer 2017). There has been some work that aims to derive a population prior to enforce the same pattern of sparsity across subjects (Varoquaux et al. 2010), but this differs from the problem here which aims to generate one connectivity graph per group. As an alternative, under the small-world hypothesis, one study aims to decompose whole-brain connectivity into decomposable smaller graphs (Varoquaux et al. 2012).

---

[1]This differs from the reweighted-$\ell_1$ minimization commonly used in compressed sensing, which typically equips a general linear model to robustly impose sparsity with very few samples (Candes, Wakin, and Boyd 2008).

**Multi-task brain studies.** Two recent studies apply multi-task learning to brain connectivity determination. MNS (Monti, Anagnostopoulos, and Montana 2015) learns population and subject-specific connectivity in brain networks, but can not effectively discern between two large classes, as is done here. Another recent model, sGGGM (Ng et al. 2013), applies sparsity in a multi-task setting to functional connectivity determination.

**Multi-task baselines.** W-SIMULE is compared to the two most-cited graphical models for multi-task learning: JGL (Danaher, Wang, and Witten 2014) and SIMONE (Chiquet, Grandvalet, and Ambroise 2011), and two more recent models with formulations closer to the one here: CLIME (Cai, Liu, and Luo 2011) and SIMULE (Wang, Singh, and Qi 2016). Additionally, all models are compared against the extremely popular graphical lasso (GLASSO) (Friedman, Hastie, and Tibshirani 2008). Since previous weighted-$\ell_1$ GGMs are not multi-task, comparisons to these models are made by lowering the parameter $\epsilon$ to eliminate W-SIMULE's multi-task component.

## 3 W-SIMULE: A Weighted-$\ell_1$, Multi-task GGM Model

The main idea behind W-SIMULE is exploiting a prior to jointly estimate connectivity for multiple groups. Figure 1 illustrates this intuition.

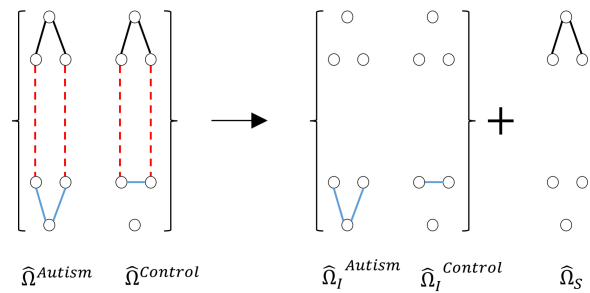

Figure 1: Toy example depicting W-SIMULE. Left shows potential edges present in the data and right shows learned edges. Long edges (red) are spatially penalized and discarded, edges that differ between groups (blue) are learned individually, and edges shared between groups (black) are learned in $\widehat{\Omega}_S$.

**Algorithm.** The problem of determining functional brain connectivity concerns using the covariance matrix ($\Sigma$) to calculate the precision matrix ($\Omega = \Sigma^{-1}$), which represents conditional correlations between brain areas. To do this, W-SIMULE takes advantage of four properties of brain-imaging data (covered in the next four subsections): (a) sparsity, (b) multi-task learning, (c) a prior, and (d) a nonparanormal assumption. In the following work, $K$ is the number of groups, $||\cdot||_1$ is the $\ell_1$ norm, $||\cdot||_\infty$ is the $\ell_\infty$ norm, $W$ is a prior matrix of positive weights, $\Sigma$ is the covariance matrix, $\Omega$ is the precision matrix, and the dot product ($\cdot$) between two matrices is their elementwise dot product.

**(a) Sparsity.** Imposing sparsity is important for interpreting brain connectivity analysis because graphs with too many connections yield very little information. In graphical models, sparsity is generally controlled with an $\ell_1$ norm. A simple example of this is the CLIME estimator (Cai, Liu, and Luo 2011), which estimates the precision matrix via constrained-$\ell_1$ minimization:

$$\widehat{\Omega} = \underset{\Omega}{\operatorname{argmin}} ||\Omega||_1 \qquad (1)$$
$$\text{subject to: } ||\Sigma\Omega - I||_\infty \leq \lambda$$

where $\lambda$ is a hyperparameter controlling the sparsity of $\Omega$.

**(b) Multi-task learning.** Multi-task learning allows the model to simultaneously estimate more than one group. For example, simply summing CLIME estimators from Eq. (1) over tasks yields a multi-task formulation:

$$\widehat{\Omega}^{(1)}, ..., \widehat{\Omega}^{(K)} = \underset{\Omega^{(i)}}{\operatorname{argmin}} \sum_i ||\Omega^{(i)}||_1 \qquad (2)$$
$$\text{subject to: } ||\Sigma^{(i)}\Omega^{(i)} - I||_\infty \leq \lambda, i = 1, ..., K.$$

where $\Omega^{(i)}$ is the precision matrix for a group $i$.

It is simple to see that multi-task learning improves performance over single-task models (Evgeniou and Pontil 2004), especially when there are few samples. However, one must choose between two multi-task modeling strategies. The first only models the differences between groups. We ignore this strategy since it does not generate full connectomes for each group, as is desired in many neural applications which require understanding whole-brain connectivity patterns. Instead, we share parameters between different groups. Mathematically, we model $\Omega^{(i)}$ as two parts:

$$\Omega^{(i)} = \Omega_I^{(i)} + \Omega_S \qquad (3)$$

where $\Omega_I^{(i)}$ is the individual precision matrix for group $i$ and $\Omega_S$ is the shared precision matrix between groups. This yields the following formulation:

$$\widehat{\Omega}_I^{(1)}, ..., \widehat{\Omega}_I^{(K)}, \widehat{\Omega}_S = \sum_i \underset{\Omega_I^{(i)}, \Omega_S}{\operatorname{argmin}} ||\Omega_I^{(i)}||_1 + \epsilon K ||\Omega_S||_1$$
$$\text{subject to: } ||\Sigma^{(i)}(\Omega_I^{(i)} + \Omega_S) - I||_\infty \leq \lambda, i = 1, ..., K. \qquad (4)$$

**(c) Weighted prior.** Over time, neuroscientists have gathered considerable knowledge regarding the spatial and anatomical priors underlying brain connectivity (*i.e.* short edges and certain anatomical regions are more likely to be connected (Watts and Strogatz 1998)). Previous studies (see Sec. 2) enforce these priors via a matrix of weights, $W$, corresponding to edges. Existing brain connectivity studies enforce a spatial prior using a weighted-$\ell_2$ norm (Ng and Abugharbieh 2011; Grosenick et al. 2011; Baldassano et al. 2012), resulting in the following penalization term: $||W \cdot \Omega||_2$. This weighted-$\ell_2$ norm effectively imposes spatial smoothness. Other studies, unrelated to brain connectivity, use a weighted-$\ell_1$ norm to enforce a prior in a graphical model: $||W \cdot \Omega||_1$. Here, we opt for the $\ell_1$ norm, as it effectively combines the prior with sparsity (Shimamura et al. 2007).

**(d) Nonparanormal Extension.** In addition to Gaussian data, W-SIMULE supports nonparanormal Gaussian data (we refer to the nonparanormal version as W-SIMULE and the Gaussian version as W-SIMULEG). This is implemented by using the Kendall correlation matrix ($\Sigma_N$) of the data matrices rather than the sample covariance matrix ($\Sigma$). This allows W-SIMULE to fit many datasets that violate the often-used Gaussian assumption by fitting a nonparanormal distribution (Liu, Lafferty, and Wasserman 2009).

**W-SIMULE: Putting it all together.** Combining the elements of sparsity, multi-task learning, a weighted prior, and a nonparanormal assumption yields the novel formulation of W-SIMULE:

$$\widehat{\Omega}_I^{(1)}, ..., \widehat{\Omega}_I^{(K)}, \widehat{\Omega}_S = \sum_i \underset{\Omega_I^{(i)}, \Omega_S}{\operatorname{argmin}} ||W \cdot \Omega_I^{(i)}||_1 + \epsilon K ||W \cdot \Omega_S||_1$$
$$\text{Subject to: } ||\Sigma_N^{(i)}(\Omega_I^{(i)} + \Omega_S) - I||_\infty \leq \lambda, i = 1, ..., K. \qquad (5)$$

W-SIMULE has three hyperparameters ($W$, $\lambda$, and $\epsilon$) that make it incredibly flexible. Using a different $W$ can enforce a different prior or change how strictly a prior is enforced. Next, changing the hyperparameter $\lambda$ controls the total sparsity of the resulting precision matrices. Finally, changing the hyperparameter $\epsilon$ allows for controlling how strictly the group penalty is imposed, *i.e.* the relative sparsities between the shared parameters and the individual parameters.

**Optimization.** Eq. (5) can be solved in parallel for each column $j$:

$$\underset{\beta^{(i)}, \beta^s}{\operatorname{argmin}} \sum_i ||W_{,j} \cdot \beta^{(i)}||_1 + \epsilon K ||W_{,j} \cdot \beta^s||_1 \qquad (6)$$

$$\text{Subject to: } ||\Sigma^{(i)}(\beta^{(i)} + \beta^s) - e_j||_\infty \leq \lambda, i = 1, \ldots, K$$

where $W_{,j}$ is the j-th column of $W$, $\beta^{(i)}$ is the j-th column of $\Omega_I^{(i)}$ of $i$-th graph (we take out the subscript $I$ in $\beta^{(i)}$ to simplify notations), and $\beta^s$ is the corresponding column in the shared part $\Omega_S$. Simplifying Eq. (6) yields

$$\operatorname*{argmin}_{\theta} ||W_{,j} \cdot \theta||_1 \quad (7)$$

Subject to: $|\mathbf{A}^{(i)}\theta - b|_\infty \leq c, \ i = 1, \ldots, K$

Where $\mathbf{A}^{(i)} = [0, \ldots, 0, \Sigma^{(i)}, 0, \ldots, 0, \frac{1}{\epsilon K}\Sigma^{(i)}]$,

$\theta = [\beta^{(1)^T}, \ldots, \beta^{(K)^T}, \epsilon K(\beta^s)^T]^T$,

$b = \mathbf{e}_j, c = \lambda$

---

**Algorithm 1** A Weighted-$\ell_1$, Multi-task Graphical Model (W-SIMULE)

---
**Input:** Data matrices $\mathbf{X}^{(i)}, \ldots, \mathbf{X}^{(K)}$, regularization hyperparameter $\lambda$, hyperparameter $\epsilon$, and linear programming solver **LP(.)**, which solves Eq. (7)
**Output:** Shared graph $\Omega_S$ and individual graphs $\Omega_I^{(1)}, \ldots, \Omega_I^{(K)}$
1: **for** $i = 1$ **to** $K$ **do**
2:     Initialize $\Sigma^{(i)}$ as the sample cov. matrix of $\mathbf{X}^{(i)}$
3:     Initialize $\Omega_I^{(i)} = \mathbf{0}_{p \times p}$
4:     Initialize $\mathbf{A}^{(i)} = [0, \ldots, 0, \Sigma^{(i)}, 0, \ldots, 0, \frac{1}{\epsilon K}\Sigma^{(i)}]$
5: **end for**
6: Initialize $\Omega_S = \mathbf{0}_{p \times p}$
7: **for** $j = 1$ **to** $p$ **do**
8:     $\theta = \mathbf{LP}(\mathbf{A}^{(i)}, b = \mathbf{e}_j, c = \lambda)$ where $i = 1, \ldots, K$
9:     **for** $i = 1$ **to** $K$ **do**
10:      $\Omega_{I,j}^{(i)} = \theta_{((i-1)p+1):ip}$
11:     **end for**
12:     $\Omega_{S,j} = \theta_{(Kp+1):(K+1)p}$
13: **end for**

---

To solve Eq. (7), we follow the simplex method (Pang, Liu, and Vanderbei 2014), which is empirically faster than the primal dual interior method (Cormen 2009). The final formulation becomes

$$\operatorname*{argmin}_{\theta^+, \theta^-} W_{,j} \cdot \theta^+ + W_{,j} \cdot \theta^-$$

Subject to :

$$\begin{pmatrix} \mathbf{A}^{(i)} & -\mathbf{A}^{(i)} \\ -\mathbf{A}^{(i)} & \mathbf{A}^{(i)} \end{pmatrix} \begin{pmatrix} \theta^+ \\ \theta^- \end{pmatrix} \leq \begin{pmatrix} c+b \\ c-b \end{pmatrix} \quad (8)$$

$$\begin{pmatrix} \theta^+ \\ \theta^- \end{pmatrix} \geq 0$$

where $\theta^+$ and $\theta^-$ refer to the positive and negative parts of $\theta$, respectively.

W-SIMULE is summarized in Algorithm 1. Following CLIME, we then apply the same symmetric operators on $\{\Omega^{(i)} = \Omega_S + \Omega_I^{(i)}\}$ obtained from Algorithm 1.

Each column of W-SIMULE can be solved in parallel. Algorithm 1 can be revised into a parallel version by modifying the "for loop" of step 7 in Algorithm 1 into a "parallel for loop" over columns. The model's convergence follows from the proven convergence of SIMULE (Wang, Singh, and Qi 2016) and the fact that the positive weights in $W$ yield a convex norm.

## 4 Experiments

This section reports experiments showing the effectiveness of W-SIMULE. It begins with the experimental setup in subsection 4.1, then details experiments showing effectiveness in subsection 4.2, and finally gives neuroscientific validation in subsection 4.3.

### 4.1 Experimental Setup

**Data.** The data examined here comes from the Autism Brain Imaging Data Exchange (ABIDE) (Di Martino et al. 2014), a publicly available resting-state fMRI dataset. The ABIDE data was released with the goal of understanding human brain connectivity and how it reflects neural disorders (Van Essen et al. 2013). The data was retrieved from the Preprocessed Connectomes Project (Craddock 2014), where preprocessing was performed using the Configurable Pipeline for the Analysis of Connectomes (CPAC) (Craddock et al. 2013) without global signal correction or bandpass filtering. After preprocessing with this pipeline, 871 individuals remain (468 diagnosed with autism). Signals for the 160 regions of interest (ROIs) in the often-used Dosenbach Atlas (Dosenbach et al. 2010) are examined.

**Priors.** To select the prior $W$, two separate spatial priors were derived from the Dosenbach atlas. The first, referred to as *anatomical$^i$*, gives each ROI one of 40 well-known, anatomic labels (*e.g.* "basal ganglia", "thalamus"). Weights take the low value $i$ if two ROIs have the same label, and the high value $10 - i$ otherwise. The second prior, referred to as *dist$^i$*, sets the weight of each edge to its spatial length, in MNI space[2], raised to the power $i$.

**Cross-validation.** Classification is performed using 3-fold cross validation, an important step in fMRI analysis (Poldrack et al. 2008), but often neglected in the literature until recently (Varoquaux et al. 2010). The subjects are randomly partitioned into 3 equal sets: a training set, a validate set, and a test set. Each model produces $\widehat{\Omega}^{\text{control}}$ and $\widehat{\Omega}^{\text{autism}}$ using the training set. Then, these graphs are fed as inputs to linear discriminant analysis (LDA), which is tuned via cross-validation on the validate set. Finally, accuracy is calculated by running LDA on the test set. Importantly, this process the ability of a method to learn the connectome's structure. The full process is performed and averaged over three folds for each model covered in Sec. 2 which produces a connectome. Notably, some of the methods (*e.g.* DPM) cannot be compared against, as they do not provide the precision matrices necessary for LDA. Other methods, (*e.g.* MNS) fail to converge when run on the dataset, as they can not handle a large number of subjects in each group.

### 4.2 Empirical Effectiveness
**Log-likelihood.** The most often-used metric for comparing graphs generated by graphical models is the log-likelihood. Here, connectomes are generated for various sparsity levels and their resulting log-likelihoods are plotted in Fig 2.[3] Unsurprisingly, as the number of total edges included in the graphs increases, the log-likelihood of the model increases. W-SIMULE ($\epsilon = 1$, $W = dist^2$), shown as the dotted red line, outperforms all of the relevant baselines, especially at low sparsities, which are most biophysically plausible. Intuitively, this suggests that W-SIMULE can find the fewest edges which explain the most observed data. This is a very useful property to neuroscientists who seek interpretable connectomes. Without the nonparanormal assumption, W-SIMULEG only outperforms other baselines at low sparsities.

---
[2] MNI space is a coordinate system used to refer to analogous points on different brains.

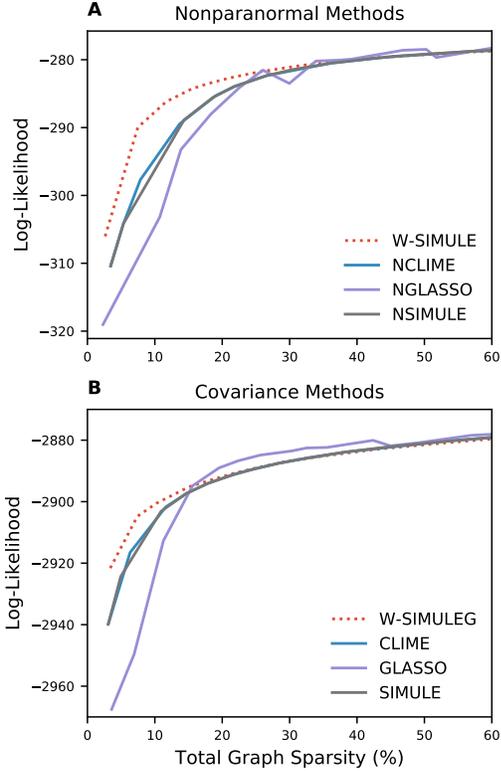

Figure 2: Model performance measured by log-likelihood. **A** and **B** show the log-likelihood versus number edges included in the model. Note that **A** and **B** are not directly comparable, as **A** uses the nonparanormal versions of the models in **B**. Both JGL and SIMONE are not shown as they yield log-likelihood values too low to be plotted on these graphs.

**Classification Accuracy.** Table 2 displays the maximum accuracy achieved for each baseline, after sweeping over hyperparameters. W-SIMULE ($\epsilon = 1$, $W = anatomical^2$, $\lambda = 0.04$) yields a classification accuracy of 58.62% between the autism and control groups, outperforming other state-of-the-art graphical models.

[3] All models were also run with intertwined covariances (covariances generated in a multi-task setting), but the results did not improve and are omitted.

Table 2: Classification accuracy obtained on the ABIDE dataset using various methods. W-SIMULE has the highest accuracy of all the values in the table.

| Method | Accuracy (%) |
| --- | --- |
| W-SIMULE | **58.62%** |
| CLIME | 46.55% |
| GLASSO | 53.71% |
| SIMULE | 57.96% |
| JGL (fused) | 56.90% |
| SIMONE | 53.71% |

**Parameter variation.** The results are fairly robust to variations of the prior (see Table 3A). The effect of changing the prior seems to have a fairly small effect on the log-likelihood of the model. This is likely because all examined priors penalize picking physically long edges, which agrees with observations from neuroscience. The *dist* prior effectively encourages the selection of short edges (see Appendix Fig A1), and the *anatomical* prior also has substantial spatial localization.

Table 3B shows that the effect of changing $\epsilon$ (the strength of the multi-task component). The log-likelihood is robust to variations in $\epsilon$ over a certain range, but can change significantly if $\epsilon$ varies drastically. The same goes for the test accuracy. Generally, as $\epsilon$ gets larger, thus increasing the importance of the shared parameters between groups, the log-likelihood increases. This emphasizes the importance of the multi-task term of W-SIMULE. Since the total number of subjects is limited, strengthening the multi-task component effectively doubles the sample size (including both classes) and allows for better picking edges.

Table 3: Variations of the prior and multi-task component yield fairly stable results.

| | A: Varying Prior ($\epsilon = 1$) | | | |
| --- | --- | --- | --- | --- |
| Prior | Log-Likelihood Sparsity=8% | Test Accuracy Sparsity=8% | Log-Likelihood Sparsity=16% | Test Accuracy Sparsity=16% |
| *No prior* | -295.98 | 0.56 | -286.17 | 0.55 |
| *dist* | -290.71 | 0.54 | -284.63 | 0.56 |
| *dist²* | -289.55 | 0.54 | -283.89 | 0.54 |
| *anatomical¹* | -290.84 | 0.55 | -283.69 | 0.56 |
| *anatomical²* | -292.14 | 0.58 | -284.72 | 0.55 |

| | B: Varying $\epsilon$ (Prior = *anatomical²*) | | | |
| --- | --- | --- | --- | --- |
| $\epsilon$ | Log-Likelihood Sparsity=8% | Test Accuracy Sparsity=8% | Log-Likelihood Sparsity=16% | Test Accuracy Sparsity=16% |
| 1.6 | -292.14 | 0.58 | -284.72 | 0.54 |
| 1.2 | -292.13 | 0.58 | -284.71 | 0.54 |
| 1.0 | -292.14 | 0.58 | -284.72 | 0.55 |
| 0.6 | -301.72 | 0.55 | -294.91 | 0.54 |

### 4.3 Neuroscientific validation

**Connectome.** W-SIMULE yields different connectomes depending on its hyperparameters. Here, Fig 3 shows the connectome which yielded the maximum accuracy in Table 2, generated using W-SIMULE ($\epsilon = 1$, $W = anatomical^2$, $\lambda = 0.04$). In order to be interpretable, only 2.5% of the possible edges were visualized (the same set of possible edges is visualized for both the autism and control groups). Note that many edges are shared between the groups, emphasizing the need for multi-task learning.

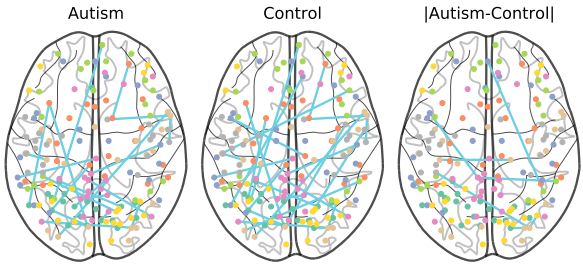

Figure 3: Sparse connectome generated by W-SIMULE using the *anatomical*$^2$ prior. The autism graph, control graph, and their difference are shown. Many edges are shared between the groups. Visualized with nilearn (Abraham et al. 2014).

**Autism-specific areas.** To analyze neural differences, we average over several connectomes. Again using $\epsilon = 1$ and $W = anatomical^2$, sweeping over different values of $\lambda$ yields connectomes at various sparsity levels. At each sparsity level, we calculate the difference between the autism and control group, as seen in Fig. 3. On average, the edges that connect to the following four areas are most affected (in decreasing order): the precuneus, the basal ganglia, the anterior cingulate cortex, and the medial frontal cortex. These results are consistent with the findings of previous brain-imaging studies. For example, some find significant under-connectivity to and from the precuneus in autistic subjects (Cherkassky et al. 2006). Together, the medial frontal cortex and anterior cingulate cortex have been linked to the the neural basis of social impairments in autism (Mundy 2003). Finally, the changes in the basal ganglia due to autism reflect changes in gait of autism patients (Rinehart et al. 2006). These results serve as validation for W-SIMULE, but note that the analysis here yields no information about the neural connections within these fairly large areas. In order to find more detailed information within one of these areas, one could run W-SIMULE using an atlas more refined than the Dosenbach atlas, which divides the brain into just 160 ROIs.

### 5 Conclusions

Here, we develop W-SIMULE, which effectively generates quantifiable, state-of-the art connectivity. W-SIMULE is highly effective and easily adapts to different domains. Connectomes generated by W-SIMULE selectively highlight connections that are important for distinguishing between autism and control groups (see Figure 2B, Table 2) and can be used to analyze the connectome (see subsection 4.3). W-SIMULE can help researchers to pinpoint the neural basis of autism or other disorders with large fMRI datasets (Milham et al. 2012; Smith et al. 2013a).

W-SIMULE has great potential for future applications. As brain-imaging datasets become more complex and include more structural data (*e.g.* MRI) coupled with functional data (*e.g.* fMRI), W-SIMULE will become increasingly important to neuroscience. This is especially true for studies with small sample sizes, such as task-specific studies, which require strong priors and multi-task learning in order to robustly determine connectivity (Real et al. 2017). As the spatial resolution of fMRI increases, spatial penalization will become more important in constructing accurate ROIs and brain connections (Craddock, Tungaraza, and Milham 2015; Thirion et al. 2014). Finally, many problems outside of neuroscience can benefit from W-SIMULE; it can utilize diverse priors to find conditional independence between nodes in any multi-task setting. Thus, W-SIMULE can be readily applied to gene-network estimation (Shimamura et al. 2007), computer vision (where physical distance could be used as a prior in images), and many other problems that currently utilize Gaussian graphical models.

### Appendix

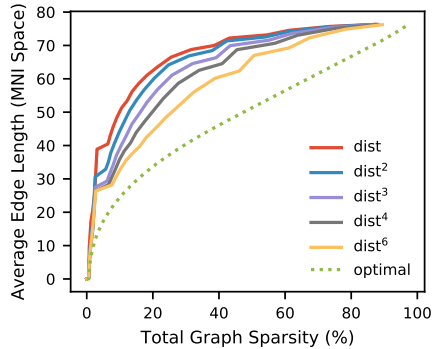

Figure A1: W-SIMULE effectively enforces the prior. As the *dist* prior is raised to a higher power (thereby increasing the spread of the weights in the matrix $W$), the prior is more strictly enforced. This results in a lower average edge length at every sparsity level. The dotted "optimal" line shows the lowest possible average edge length as a function of graph sparsity. Here, W-SIMULE uses $\epsilon = 1$ and $\lambda$ is varied to yield different sparsities.

Table A1: Classification accuracy obtained on ABIDE dataset by various studies. In general, classifiers significantly improve over randomness (50%). These studies are not directly comparable to W-SIMULE. Most take drastically different approaches and do not provide a full, interpretable connectome as W-SIMULE does. Instead, many classify without learning the connectivity structure (*e.g.* with a neural network). The accuracy score also does not consider sparsity, domain adaptivity, or efficiency. Preprocessing, training, and validation schemes varies between the studies. Smaller subsets of the data are generally able to achieve better performance.

| Study | Method | Total Subjects | Autism Subjects | Control Subjects | Accuracy (%) |
|---|---|---|---|---|---|
| Ghiassian et al. 2013 | MRMR | 1111 | 538 | 573 | 63 |
| Nielsen et al. 2013 | GLM | 964 | 447 | 517 | 60 |
| Haar et al. 2014 | LDA/QDA | 906 | 453 | 453 | ∼50 |
| W-SIMULE | LDA | 871 | 403 | 468 | 58.6 |
| Parisot et al. 2017 | Graph CNN | 871 | 403 | 468 | 69.5 |
| Abraham et al. 2017 | See paper | 871 | 403 | 468 | 66.8 |
| Iidaka 2015 | PNN | 640 | 328 | 312 | 90 |
| Haar et al. 2014 | LDA/QDA | 590 | 295 | 295 | 60 |
| Chen et al. 2015 | RF | 252 | 126 | 126 | 91 |
| Chen et al. 2016 | SVM | 240 | 112 | 128 | 79 |
| Plitt, Barnes, and Martin 2015 | L2LR | 178 | 89 | 89 | 71 |